\documentclass[sn-chicago]{sn-jnl}

\usepackage{graphicx,setspace,lscape,longtable}
\usepackage{natbib,epsfig,pdfpages}
\usepackage{rotating}
\usepackage{float}
\usepackage{bm}
\usepackage[normalem]{ulem}
\usepackage{CJK}
\usepackage{subfigure}
\usepackage{array,booktabs}
\usepackage{algorithm}
\usepackage{algpseudocode}
\usepackage{setspace}
\usepackage{enumerate}
\usepackage{paralist}
\usepackage{fancyhdr}
\usepackage{geometry}

\theoremstyle{thmstyleone}%
\theoremstyle{thmstyletwo}%

\theoremstyle{thmstylethree}%
\begin{document}

\title[ICIM]{A Constrained Spatial Autoregressive Model for Interval-valued data}


\author{\fnm{Tingting} \sur{Huang}}\email{tingth@cueb.edu.cn}

\affil{\orgdiv{School of Statistics}, \orgname{Capital University of Economics and Business}, \orgaddress{\city{Beijing}, \postcode{100070}, \country{China}}}


\abstract{ Interval-valued data receives much attention due to its wide applications in the fields of finance, econometrics, meteorology and medicine. However, most regression models developed for interval-valued data assume observations are mutually independent, not adapted to the scenario that individuals are spatially correlated. We propose a new linear model to accommodate to areal-type spatial dependency existed in interval-valued data. Specifically, spatial correlation among centers of responses are considered. To improve the new model's prediction accuracy, we add three inequality constrains. Parameters are obtained by an algorithm combining grid search technique and the constrained least squares method. Numerical experiments are designed to examine prediction performances of the proposed model. We also employ a weather dataset to demonstrate usefulness of our model.}




\maketitle

\section{Introduction}\label{sec1}

Rapid development of information technology makes collect and store data at a low cost, resulting in huge, heterogeneous, imprecise, and structured data sets to be analyzed. Particularly, we may deal with a type of data, whose units of information are intervals consisted of lower and upper bounds. The new-type data arise in various contexts.

For example, if our question of interest involves community  features, then there is a need to aggregate classical numerical data table of individuals into a smaller data sheet of groups (\cite{billard_statistics_2003, wei_interval-valued_2017}). The requirement of data confidentiality or lower computation complexity can produce such intervals as well. We summarize these sources of intervals as information gathering. What's more, some real data endowed with variability are naturally represented by an interval, for instance, daily temperature with the maximum and the minimum values, stock prices at the opening and the closing (\cite{lim_interval-valued_2016, sun_interval-valued_2022}). Imprecise observations and uncertainty are the third way causing intervals, which takes advantages of intervals to represent error measurements or error estimations of possible values.

We call this complex data interval-valued data. Modeling real data by intervals, we can better take fluctuations, oscillations of variables into consideration, and understand things from a more comprehensive view (\cite{wang_cipca_2012}). Since \cite{diday1988symbolic} proposed the concept of symbolic data analysis, interval-valued data, as the most common type of symbolic data, has been extensively studied.

Regression analysis as one of the most popular approaches in statistics, exploring how a dependent variable changes as one or more covariates are changed, has been extended to investigate interval-valued data as well. Currently, discussing the relation between interval-valued responses and interval-valued predictors receives the most attention. As the data units are comprised of two bounds, not just single points, applying classical linear models to interval-valued data should be careful. Two paths of  modelling interval-valued data in regressions have been developed(\cite{sun_linear_2016}). One presumes interval-valued data sit in a linear space, endowed with certain addition and scalar multiplication operations. Thus the aim is find to find a function of the interval-valued explanatory variables so that it is "closest" to the outcomes. The other treats intervals as a two-dimensional vector, then establishing two separate regressions for lower and upper bounds, or centers and ranges. Although the second method may lose some geometric features of interval-valued data, it brings simpler mathematical operations and accordingly more flexibility.

We mainly focus on literature into the second method. \cite{bock_symbolic_2002} constructs two linear models for the lower and upper bounds of the response interval (the MinMax method). \cite{lima_neto_centre_2008} on the other way, suggests to fit mid-point and semi-length of the interval-valued dependent variable instead of the bounds (the center and range method, CRM).  \cite{lima_neto_constrained_2010} then further incorporates a nonnegative constraint to the coefficient of range model so as to ensure the predicted lower boundaries smaller than the upper ones (the constrained center and range method, CCRM). To improve model's prediction accuracy, \cite{hao_constrained_2017} adds two other inequality restrictions to the constrained center and range model so that predicted intervals and true values have overlapping areas. Besides, variants of interval-valued linear model have also been extensively studied. \cite{fagundes_interval_2014} shows an interval kernel regression. \cite{lima_neto_nonlinear_2017} introduces nonlinear interval-valued regression. \cite{lim_interval-valued_2016} proposes nonparametric additive interval-valued regression. \cite{lima_neto_exponential-type_2018} presents a robust model for interval-valued data by penalizing outliers through exponential kernel functions. \cite{yang_interval-valued_2019} introduces artificial neural network techniques to interval-valued data prediction. More recently, \cite{de_carvalho_clusterwise_2021} combines k-means algorithm with linear and nonlinear regressions and raises clusterwise nonlinear regression for interval-valued data. \cite{xu_bivariate_2022} extends the center and range method to the Bayesian framework. Notably, all these investigations are conducted with the assumption that observations are mutually independent.

However, we may encounter real data that are spatially correlated in practice. Like in weather data, seasonal precipitations of a city are often high if it's neighbor cities have abundant rainfall. And in unemployment data, unemployment rates of different districts in the same city are spatially agglomerate. Housing price data and stock price exhibit similar patterns (\cite{Topa2001Social,anselin1998spatial,Lesage2009Introduction}). An vital and common characteristic of these correlated data is that spatial dependence between individuals $i$ and $i^\prime$ is determined by distance $d_{ii^\prime}$ between them. The closer two spatial units are (the smaller $d_{ii^\prime}$), the greater the spatial dependency. Causes of such correlation are mainly from human spatial behavior, spatial interaction and spatial heterogeneity. In applications, $d_{ii^\prime}$ can be a more general distance, such as social distance, policy distance and economic distance (\cite{19711801501}). Areal data (lattice data) is one kind of this spatially correlated data. It arises when a fixed domain is carved up by a finite number of subareas. Resulting in each observation of areal data is aggregated outcomes in the same subregion. Lattice data receive much attention in the fields of econometrics and spatial statistics. We focus on modelling interval-valued data with areal-type spatial dependence in this research.

\begin{figure}[!h]
  \centering
    \subfigure{
    \resizebox{3.72cm}{3.3cm}{\includegraphics{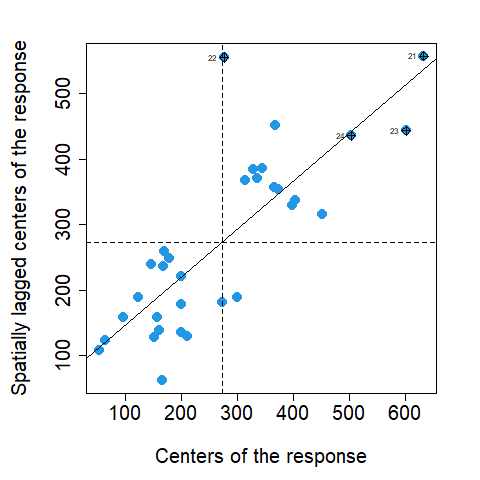}}}
   \subfigure{
  \resizebox{3.72cm}{3.3cm}{\includegraphics{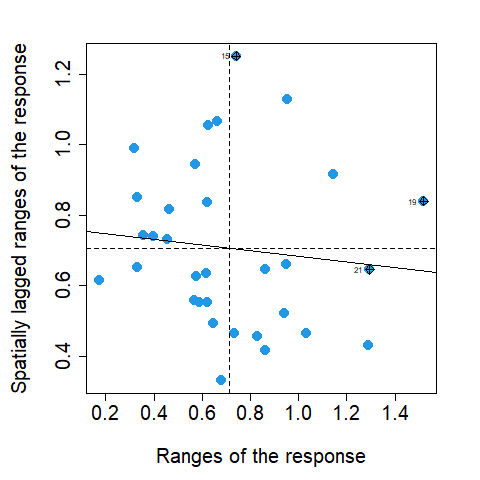}}}
  \subfigure{
  \resizebox{3.92cm}{3.3cm}{\includegraphics{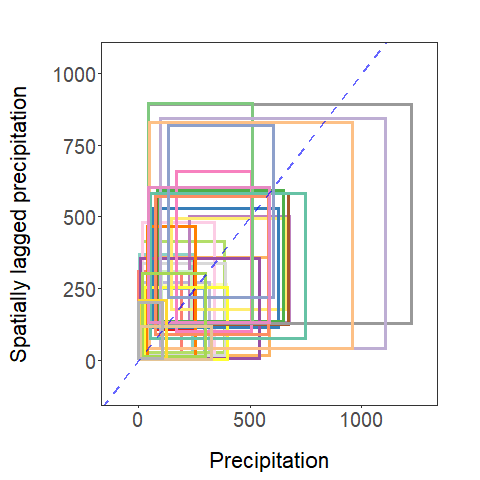}}}
  \caption{The Moran`I Scatter plot of the centers (left) and the ranges (middle), the predicted precipitation versus the true values (right).} \label{fig1-1}
\end{figure}

To better illustrate our motivation, we present a practical example. To investigate how temperature affects precipitation in major cities of China, monthly weather data from China Meteorological Yearbook of $2019$ were collected. As temperature data are naturally interval-valued data, interval-valued linear regressions are employed to analyze the problem. In data preprocessing, the single-valued monthly rainfall and monthly temperature for each city were aggregate into intervals. Then to see whether the interval-valued outcomes are spatially dependent, the Moran'I statistic was utilized. We tested spatial correlation among centers and radii of the interval-valued precipitations, separately. The values of the statistics are 0.675 for centers with p-value equaling $0.00001$, and $-0.081$ for ranges with p-value equaling $0.6149$. Obviously, mid-points of the interval-valued response are correlated, and semi-lengths are not. If apply exiting interval-valued linear regressions to model the problem, the results obtained would be misleading as these models mostly assume observations are mutually independent. Therefore, there are reality needs to model interval-valued data with spatial correlation. Figure \ref{fig1-1} (left and middle) shows Moran'I scatter plots of the centers and the ranges. Figure \ref{fig1-1} (right) displays Moran's I scatter plot of the interval-valued response, where the spatially lagged intervals on the vertical axis are aggregated from single-valued lagged precipitations. The spatial weight matrix $\bm{W}$ used here will be illustrated at length in Section \ref{sec5}.

\vspace{-0.4cm}
\vspace{-0.45cm}

When only the ordinary single-valued data are invloved, spatial linear models, including spatial autoregressive model (SAR), spatial error model (SEM) and spatial Durbin model (SDM) are most commonly employed to analyze areal data. Through the utility of spatial weight matrix $\bm{W}$ and spatial lag parameter $\rho$, spatial models incorporates spatial effects into classic linear regressions. Estimation methods for these regressions cover quasi-maximum likelihood estimation method (\cite{Lee2004Asymptotic}), generalized moment estimators (\cite{Lee2007GMM}) and the Markov Chain Monte Carlo (MCMC) method (\cite{Lesage2009Introduction}). Their applications range from stock markets to social networks (\cite{ma_naive_2020,zhang_spatial_2019}). As the SAR model is the most popular spatial model, we capitalize on it to struct our new model.

In this article, we propose a constrained spatial autoregressive model for interval-valued data (ICSM), which aims to predict interval-valued outcomes in the presence of areal-type spatial dependency. In particular, the CRM (center and range method)
is utilized to handle inter-valued data. And we considered spatial correlation among centers of responses, employing the SAR model to form regression. We also take prediction accuracy into account by adding three inequality constrains in the model, motivated by \cite{hao_constrained_2017}. The parameters are obtained by the combination a grid search technique and the constrained least squares method. We also conduct numerical experiments to examine prediction performances of the proposed model. We find when spatial correlation is great and signal to noise is high, our model behaves better than models without spatial term and constrains. The weather data mentioned earlier is thoroughly analyzed by the new model.

We note that literatures related to spatial linear models of interval-valued data are relatively scanty. \cite{gonzalez-rivera_vector_2016} discusses vector autoregressive moving average model for interval-valued data. \cite{sun_threshold_2018} proposes a new class of threshold autoregressive interval models. Although these investigations have considered correlation between intervals, they are mainly designed for interval-valued time series data, not  interval-valued data with lattice-type spatial dependency.

The paper is organized as follows. Section \ref{sec2} introduces the spatial autoregressive model and our proposed  ICSM. Section \ref{sec3} describes the algorithm to obtain the parameters. Numerical experiments are conducted in Section \ref{sec4} to examine performances of the new model. Real data analysis are arranged in Section \ref{sec5}. We end up the article with a conclusion.

\section{Model Specification}\label{sec2}
\subsection{Spatial autoregressive (SAR) model}
\cite{Ord1975Estimation} proposed the spatial autoregressive (SAR) model, modeling spatial effects in the areal data. Suppose there are $n$ spatial units on a lattice, and each individual occupies a position. We observe $\{(x_i ,y_i )\}_{i=1}^n$ from these entities. Denote $\bm{x}=\big(x_1,x_2 ,\cdots,x_n \big)^\top$ and $\bm{y}=(y_1,y_2,\cdots,y_n)^\top$, the SAR model is formulated as
$$
\bm{y}=\rho\bm{W}\bm{y}+\bm{x}\beta+\bm{\epsilon},~~~~\epsilon_i\sim N(0,\sigma^2)$$
where $\bm{W}=(w_{ii^\prime})_{n\times n}$ is a spatial weight matrix, with each entry $w_{ii^\prime}$ representing the weight between unit $i$ and unit $i^\prime$, $\rho\in(-1,1)$ is the spatial lag parameter to be estimated, $\bm{\epsilon}=(\epsilon_1,\cdots,\epsilon_n)^{\top}$ is the noise term independent of $\bm{x}$ and i.i.d. following a normal distribution. $\beta$ is the scalar coefficient. Here, $\bm{W}$ quantifies spatial structure of the lattice, and the spatial lag term $\rho\bm{Wy}$ models spatial effects. The greater value $\rho$ takes, the stronger spatial impacts are imposed on $\bm{y}$.

It should be emphasized that generally $\bm{W}$ is exogenous and pre-specified, where $w_{ii^\prime}$ is determined by the contiguity or distance between $i$ and $i^\prime$ in different spatial scenarios. Rook matrix is among the most common spatial weight matrices, where
$$w_{ii^\prime}=\begin{cases}
1,  & i ~and ~i^\prime ~share ~an~edge \\
0, & otherwise.
\end{cases}$$
Bishop matrix is built in similar way by setting $w_{ii^\prime}=1$ when $i$ and $i^\prime$ have the same vertex. The key point to construct such matrix is knowing the map indicating spatial arrangement of the individuals. In case that geographical locations of spatial units are given, $w_{ii^\prime}$ can be functions of distance $d_{ii^\prime}$ between the two units, for example,
$$w_{ii^\prime}=\begin{cases}
\frac{1}{d_{ii^\prime}},  & d_{ii^\prime}\leq d_0 \\
0, & d_{ii^\prime}>d_0
\end{cases}$$
where $d_0$ is a threshold distance. More general spatial weights can be derived from networks. For instance under a social network setting, $w_{ii^\prime}=1$ if person $i$ and $i^\prime$ are friends. For more information of $\bm{W}$, refer to \cite{19711801501} and \cite{anselin1998spatial}. In general, $\bm{W}$ is row-normalized so that summation of the row elements is unity. Entries on the diagonal of $\bm{W}$ are zeros.

\subsection{Constrained spatial autoregressive model for interval-valued data (ICSM)}\label{sec3}
Following \cite{Jenish2009Central}, we presume the spatial process is located on an unevenly spaced lattice $L\in R^n~ (n\geq1)$, and all units on $L$ are endowed with positions ($L$ is crucial for establishing the weight matrix $\bm{W}$).

Denote an interval-valued variable as $\bm{w}^s =[w^l,w^u]$, where $\bm{w}^s \in \mathcal{S}=\{[a,b]~\|~ a,b \in \mathbb{R},a\leq b\}$. Here, we observe $\{\bm{y}_i^s, \bm{x}_i^s\}_{i=1}^n$ from lattice $L$, and $\bm{y}_i^s=[y_i^l, y_i^u],~\bm{x}_i^s=[x_i^l,x_i^u]$. To deal with the interval-valued data, we employ the CRM method (\cite{lima_neto_centre_2008}). Specifically, represent midpoint and radius of $\bm{w}^s$ by $w^c$ and $w^r$, and write
$\bm{y}^c=(y_1^c,\cdots,y_n^c)^\top,~\bm{y}^r=(y_1^r,\cdots,y_n^r)^\top,~\bm{x}=(\bm{x}^c,\bm{x}^r)=((x_1^c,
\cdots,x_n^c)^\top,(x_1^r,\cdots,x_n^r)^\top)$, then the constrained spatial autoregressive interval-valued model (ICSM) is formulated as
\newlength{\widest}
\settowidth{\widest}{$\bm{y}^c-\bm{y}^r \leq (\bm{I}_n-\rho\bm{W})^{-1}\bm{x\beta}^c+\bm{x\beta}^r$}
\begin{equation}\label{cscrm}
  \begin{aligned}
     \begin{cases}
      \makebox[\widest][l]{$\bm{y}^c =  \rho\bm{Wy}^c+\bm{x}\bm{\beta}^c+\bm{\epsilon}^c$} &\\
      \bm{y}^r = \bm{x}\bm{\beta}^r+\bm{\epsilon}^r &
      \end{cases}\\
      s.t.
     \begin{cases}
     \makebox[\widest][l]{$\bm{y}^c-\bm{y}^r \leq (\bm{I}_n-\rho\bm{W})^{-1}\cdot \bm{x\beta}^c+\bm{x\beta}^r$} &\\
     \bm{y}^c+\bm{y}^r \geq (\bm{I}_n-\rho\bm{W})^{-1}\cdot\bm{x\beta}^c-\bm{x\beta}^r  &\\
     \bm{x\beta}^r \geq \bm{0}&
    \end{cases}
\end{aligned}
\end{equation}
where $\bm{W}$ is the spatial weight matrix, $\rho\in (-1,1)$ is the spatial lag parameter, $\bm{\beta}^c=(\beta_1^c,\beta_2^c)$ and $\bm{\beta}^r=(\beta_1^r,\beta_2^r)$ are the coefficients, and $\bm{\epsilon}^c=(\epsilon^c_1, \cdots, \epsilon^c_n)^\top, \bm{\epsilon}^r=(\epsilon^r_1, \cdots, \epsilon^r_n)^\top$ are the error terms with $E(\bm{\epsilon}^c)=\bm{0}, Var(\bm{\epsilon}^c)=\sigma^2\bm{I}_n, E(\bm{\epsilon}^r)=\bm{0}, Var(\bm{\epsilon}^r)=\sigma^2\bm{I}_n$.

Note that compared to precisely forecast semi-length $\bm{y}^r$, accurately predicting mid-point $\bm{y}^c$ would contribute more to obtain a good predictor of $\bm{y}^s$. Therefore, considering simplicity and efficiency of the new model, we give priority to modeling spatial correlation of $\bm{y}^c$ by the SAR regression, fitting $\bm{y}^r$ by ordinary linear regression. We also considered effects from both centers and ranges of $\bm{x}^s$.

Inspired by \cite{hao_constrained_2017}, to improve prediction accuracy of the model, we add three inequality constraints to the interval-valued spatial regression, where the first two ensure predicted intervals have overlapping areas with the true values, and the last guarantees predicted value of $\bm{y}^r$ is positive. Concretely, in the first inequality, the forecasted upper bound of $\bm{y}^s$ should be greater than the observed lower bound of $\bm{y}^s$; and in the second, the forecasted lower bound of $\bm{y}^s$ should be smaller than the true upper bound of $\bm{y}^s$.

In model (\ref{cscrm}), $\rho$ quantifies average spatial effects among $\bm{y}^s$'s central positions. In particular, when $\rho=0$ the ICSM degenerates into the ICM (\cite{hao_constrained_2017}). Thus (\ref{cscrm}) is a more general model.

\section{Parameter estimation}\label{sec3}

Estimating $\rho,~\bm{\beta}^c$ and $\bm{\beta}^r$ of (\ref{cscrm}) is not straightforward due to the presence of spatial lag term $\rho\bm{Wy}^c$ and the constrains. Regarding $\rho$, commonly used methods to obtain it's estimator include maximum likelihood estimation method (\cite{Lee2004Asymptotic}), generalized moment estimators (\cite{Lee2007GMM}) and Markov Chain Monte Carlo method (\cite{Lesage2009Introduction}). However, these approaches can not adapt to the situation that model parameters are restricted by inequalities. We put forward an algorithm that combines a grid search method with the constrained least squares to obtain estimators.

Firstly, as $\rho$ is a scalar number taking values from $-1$ to $1$, we can give a sufficiently fine grid of $m$ values for $\rho$. For example, take step size by $0.01$, then $m=201$ and $(\rho_1,\rho_2,\rho_3,\cdots,\rho_{109},\rho_{200},\rho_{201})=(-1,-0.99,-0.98,\cdots,0.98,0.99,1)$.

Secondly, for each value of $\rho$, we solve a constrained least squares optimization problem. Provided value of $\rho_j ~(j=1,\cdots,m)$, we can calculate $$\bm{A}_j=(\bm{I}_n-\rho_j\bm{W}).$$
Then model (\ref{cscrm}) turns into
\begin{equation*}
\left\{
\begin{array}{ll}
  \bm{y}^c = \bm{A}_j^{-1}\cdot (\bm{x}\bm{\beta}^c+\bm{\epsilon}^c)  \\
 \bm{y}^r = \bm{x}\bm{\beta}^r+\bm{\epsilon}^r \\
\end{array}
\right.
 \quad  s.t.  \quad \left\{\begin{array}{ll}
\bm{y}^c-\bm{y}^r \leq \bm{A}_j^{-1}\cdot \bm{x\beta}^c+\bm{x\beta}^r\\
\bm{y}^c+\bm{y}^r \geq \bm{A}_j^{-1}\cdot\bm{x\beta}^c-\bm{x\beta}^r \\
 \bm{x\beta}^r \geq \bm{0} \\
\end{array}\right.
\end{equation*}
Now that $\bm{A}_j$ is known, we can employ least squares method to obtain $\hat{\bm{\beta}}^c$ and $\hat{\bm{\beta}}^r$. The sum of the squares of deviations is given by
\begin{equation}\label{opt}
\begin{split}
&\min_{\bm{\beta}^c,\bm{\beta}^c} \, ~~(\bm{\epsilon}_j^c)^\top\bm{\epsilon}_j^c+(\bm{\epsilon}_j^r)^\top\bm{\epsilon}_j^r\\
&s.t.\quad  \left\{\begin{array}{ll}
-\bm{A}_j^{-1}\cdot \bm{x\beta}^c-\bm{x\beta}^r \leq -\bm{y}^c+\bm{y}^r\\
\bm{A}_j^{-1}\cdot\bm{x\beta}^c-\bm{x\beta}^r \leq \bm{y}^c+\bm{y}^r\\
-\bm{x\beta}^r \leq \bm{0} \\
\end{array}\right.
\end{split}
\end{equation}
where $\bm{\epsilon}_j^c=\bm{A}_j\cdot \bm{y}^c- \bm{x\beta}^c$ and $\bm{\epsilon}_j^r=\bm{y}^r- \bm{x\beta}^r$.

Once the solution for (\ref{opt}) is found for all candidates $\rho_j$, we select the $\rho_j$ with the least sum of square residuals as the $\hat{\rho}$. Accordingly, if the $j$th $\rho$ is the optimum $\rho$, then the $j$th estimated $\hat{\bm{\beta}}^c$ and $\hat{\bm{\beta}}^r$ are estimators of $\bm{\beta}^c$ and $\bm{\beta}^r$. We summarize the whole process in Algorithm 1.

\begin{algorithm}[htbp]
  \caption{Main steps of the estimation procedure}
  \begin{algorithmic}[1]
  \Require the dataset $\bm{y}^c,\bm{y}^r,\bm{x}^c,\bm{x}^r$ and the spatial matrix $\bm{W}$.
  \State Generate a sequence of $\rho$ from $-1$ to $1$ with step size $0.01$, that is $(\rho_1,\cdots,\rho_{200},\rho_{201})=(-1,\cdots,0.99,1)$.
  \For {$k=1$ to $201$}
    \State compute $\bm{A}=\bm{I}_n-\rho_k\bm{W}$
    \State optimize $$~~~\min \|\bm{Y}-\bm{Z\beta}\|^2 ~s.t. ~\bm{G\beta}\leq \bm{h}~~~~$$
      where $\bm{G}=\left(
              \begin{array}{cc}
                -\bm{A}^{-1}\bm{X} &  -\bm{X} \\
                 \bm{A}^{-1}\bm{X}& -\bm{X} \\
                 \bm{0}_{n\times 3}&  -\bm{X} \\
              \end{array}
            \right)
    ,\bm{h}=\left(
               \begin{array}{c}
                 \bm{y}^r-\bm{y}^c \\
                 \bm{y}^c+\bm{y}^r \\
                 \bm{0}_{n\times 1} \\
               \end{array}
             \right)
    ,\bm{Z}=\left(
               \begin{array}{cc}
                 \bm{X} & \bm{0}_{n\times 3} \\
                 \bm{0}_{n\times 3} & \bm{X}  \\
               \end{array}
             \right)
   ,\bm{Y}=\left(
                    \begin{array}{c}
                      \bm{Ay}^c \\
                      \bm{y}^r \\
                    \end{array}
                  \right)
   ,~\bm{\beta}=\left(
                  \begin{array}{c}
                    \bm{\beta}_1 \\
                    \bm{\beta}_2 \\
                  \end{array}
                \right)
    ,~\bm{X}=\left(
               \begin{array}{ccc}
                 \bm{I}_n & \bm{x}^c & \bm{x}^r \\
               \end{array}
             \right)
    ,~\bm{\beta}_1=\left(
                     \begin{array}{c}
                       \beta_0^c \\
                       \bm{\beta}^c \\
                     \end{array}
                   \right)
    ,~\bm{\beta}_2=\left(
                     \begin{array}{c}
                       \beta_0^r \\
                       \bm{\beta}^r \\
                     \end{array}
                   \right)$.
  \EndFor
  \State choose $\rho_j$ with the least $\|\bm{Y}-\bm{Z\beta}\|^2$, corresponding $\bm{\beta}$ is the estimated coefficient.
  \Ensure $\rho_j$ and the $j$th $\bm{\beta}$
  \end{algorithmic}
\end{algorithm}

\section{simulation study}\label{sec4}
In this section, numerical experiments are conducted to show the superiority of our new method. Firstly, we introduce the data generating process. Secondly, we briefly explain how unobserved individuals are predicted under a spatial scenario. Two competing regression models are also considered. Lastly, we present several criterions to measure model performances and display the experimental results. 

\subsection{data generating process}

For the spatial scenarios, we take two spatial weight matrices $\bm{W}$ into account, one is the commonly used rook matrix, and the other is the block matrix. These two matrices imply two cases of spatial structures, where the first corresponds to sparse networks, and the second corresponds to dense ones. More specifically, the spatial units under a rook scenario have a small number of neighbors, while the individuals under a block setting have a relatively greater number of neighbors. Here are details of generating the two matrices.
\begin{itemize}
  \item Rook matrix assigns $w_{ii^\prime}=1$ if units $i$ and $i^\prime$ share an edge, and  $w_{ii^\prime}=0$ otherwise. We assume $n$ agents are randomly located on a regular square lattice with $R$ rows and $T$ columns, where each agent occupies a cell. Then we have sample size $n=R\times T$. In this context, the agents have $4$ neighbors at most, and $2$ neighbors at least, given that they may be in the inner field or in the corner of the square grid. We set $n = \{10 × 12, 12 × 20, 20 × 25\}=\{120,240,500\}$.

  \item Block matrix was first discussed by \cite{Case1991Spatial}, where there are $D$ number of districts, $M$ members in each district. Specifically, each unit $i$ in the block matrix has $m-1$ neighbors with equal weights, i.e., we set $w_{ii^\prime}=\frac{1}{m-1}$ if $i$ and $i^\prime$ are in the same district, $w_{ii^\prime}=0$ otherwise. We take $n=D\times M=\{20\times6, 20\times 12, 25\times 20\}=\{120,240,500\}$

\end{itemize}

The numerical data are generated using the following model
\begin{equation*}\label{numodel}
\left\{
\begin{array}{ll}
  \bm{y}^c &= (\bm{I}_n-\rho\bm{W})^{-1} \cdot  (\bm{x}^c\beta_1^c+\bm{x}^r\beta_2^c+\bm{\epsilon}^c)  \\
  \bm{y}^r &=\bm{x}^c\beta_1^r+\bm{x}^r\beta_2^r+\bm{\epsilon}^r
\end{array}
\right.
\end{equation*}

\begin{enumerate}[1)]
  \item Regarding $\rho$, we consider three cases.
    \begin{enumerate}[i.]
        \item $\rho=0$, where spatial dependency does not present. The ICSM degenerates into the CCRJM of \cite{hao_constrained_2017}.
        \item $\rho=0.4$, where network effects are moderate.
      \item $\rho=0.8$, where network effects are strong.
      \end{enumerate}

  \item Centers and ranges of the interval covariates are generated by uniform distribution, i.e. $$\bm{x}^c\stackrel{i.i.d}{\sim} U(0,150), ~~~~~\bm{x}^r\stackrel{i.i.d}{\sim} U(5,8)$$
  \item Coefficients of the covariates $\beta_1^c, \beta_2^r$  are generated by uniform distribution, as well
   $$\beta_1^c\sim U(-2.5,-2), ~~~~~~\beta_2^r \sim  U(2.5,5),~~~~~~\beta_2^c=1,~~~~\beta_1^r=0.1$$

  \item To see how the prediction performance is influenced by different levels of noises of the center, we let
      \begin{enumerate}[i.]
        \item $~~~\epsilon_i^c\sim U(0,11),~~~~ \epsilon_r \sim U(0,5)$.
        \item $~~~\epsilon_i^c\sim U(0,18), ~~~~\epsilon_r \sim U(0,5)$.
      \end{enumerate}
\end{enumerate}

\subsection{Predicting formula and comparative models}

Note that in a spatial context, predicting $Y$ of an out-of-sample should be careful, as adding new observations will change the existing spatial structure, i.e., $\bm{W}$ is different. \cite{Goulard2017} studied the problem. They discussed out-of-sample prediction for the SAR model and concluded the ``BP'' predictor behaves well among other predictors. Denote in-sample observations by $(X_s, Y_s)$ and out-of-sample covariate by $X_o$, which are all known. Then the unknown out-of-sample response $Y_o$ can be calculated by ``BP" predictor
$$\hat{Y}_o^{BP}=\hat{Y}_o^{TC}-Q_o^{-1}Q_{os}(Y_s-\hat{Y}_s^{TC})$$
$$\hat{Y}^{TC}=(I_n-\hat{\rho}W)^{-1}X\hat{\beta}=\begin{pmatrix}
\hat{Y}_s^{TC} \\
\hat{Y}_o^{TC}
\end{pmatrix}$$
$$Q=\frac{1}{\hat{\sigma}^2}(I_n-\hat{\rho}(W^{\top}+W)+\hat{\rho}^2W^\top W)=\begin{pmatrix}
Q_s & Q_{so} \\
Q_{os} & Q_o
\end{pmatrix}$$
where $\hat{\sigma}^2$ is mean square error of the fitting model. In our simulated data set, we randomly select $\frac{9}{10}n$ individuals as the learning set, and leave the remaining $\frac{1}{10}n$ units as the test set.

We compare the ICSM's predictive power with other two models': the constrained interval-valued model without spatial dependency (ICM), and the spatial interval-valued model without constrains (ISM), which can be formulated as

\settowidth{\widest}{$-(\bm{x}^c \beta_1^c+ \bm{x}^r \beta_2^c)-(\bm{x}^c \beta_1^r+ \bm{x}^r \beta_2^r) \leq -\bm{y}^c+\bm{y}^r$}
\begin{itemize}
  \item The ICM (equivalent to the CCRJM by \cite{hao_constrained_2017})
\begin{equation}\label{cscrm}
  \begin{aligned}
     \begin{cases}
          \makebox[\widest][l]{$\bm{y}^c = \bm{x}^c \beta_1^c+ \bm{x}^r \beta_2^c +\bm{\epsilon}^c$ } &\\
      \bm{y}^r = \bm{x}^c \beta_1^r+ \bm{x}^r \beta_2^r +\bm{\epsilon}^r
      \end{cases}\\
      s.t.
     \begin{cases}
     \makebox[\widest][l]{$-(\bm{x}^c \beta_1^c+ \bm{x}^r \beta_2^c)-(\bm{x}^c \beta_1^r+ \bm{x}^r \beta_2^r) \leq -\bm{y}^c+\bm{y}^r$ } &\\
     \bm{x}^c \beta_1^c+ \bm{x}^r \beta_2^c- (\bm{x}^c \beta_1^r+ \bm{x}^r \beta_2^r) \leq \bm{y}^c+\bm{y}^r  &\\
     -( \bm{x}^c \beta_1^r+ \bm{x}^r \beta_2^r) \leq \bm{0}. &
    \end{cases}
\end{aligned}
\end{equation}

\item The ISM
\begin{equation}
\left\{
\begin{array}{ll}
  \bm{y}^c &=  \rho\bm{Wy}^c+\bm{x}^c \beta_1^c+ \bm{x}^r \beta_2^c+\bm{\epsilon}^c  \\
  \bm{y}^r  &=  \bm{x}^c \beta_1^r+ \bm{x}^r \beta_2^r+\bm{\epsilon}^r.
\end{array}
\right.
\end{equation}
\end{itemize}

\subsection{measurements and comparison results}
A number of criterion classes have been used to evaluate the performance of the interval linear regressions.  \cite{lim_interval-valued_2016} calculates root mean square error (RMSE) of the lower and upper bound of the predictive interval, i.e., RMSE$_l$ and RMSE$_u$. \cite{hojati2005} employs the accuracy rate (AR) to access similarity of two intervals based on the percentage of overlapping areas. We use these three criteria in our simulation, which are defined as
$$RMSE_l=\sqrt{\frac{1}{n}\sum_{i=1}^n(y_{i}^l-\hat{y}_i^l)^2}$$
$$RMSE_u=\sqrt{\frac{1}{n}\sum_{i=1}^n(y_{i}^u-\hat{y}_i^u)^2},
$$
$$
AR=\frac{1}{n}\sum_{i=1}^{n}\frac{\mu(\bm{y}_i\cap\hat{\bm{y}}_i)}{\mu(\bm{y}_i\cup\hat{\bm{y}}_i)},
$$
where $\mu(\bm{y}_i\cap\hat{\bm{y}}_i)$ and $\mu(\bm{y}_i\cup\hat{\bm{y}}_i)$ are volumes of intersection and union of $\bm{y}_i$ and $\hat{\bm{y}_i}$, respectively.

Note that $AR$ is an average measurement of the overlapping areas. It may happen that the value of $AR$ is great, but some $\hat{\bm{y}_i}$ are mispredicted (the corresponding $\mu(\bm{y}_i\cap\hat{\bm{y}}_i)$ equals $0$). To know how many individuals are wrongly forecasted in a data set, we add a new indicator $N_d$, which is the number of disjoint intervals, i.e.,
$$
N_{d} = \#~\{i ~\|~\mu(\bm{y}_i\cap\hat{\bm{y}}_i)=0,~i=1,\cdots,n\}.
$$

The whole process was repeated $75$ times in the experiment. We computed the average $MSE_l, MSE_u, AR, N_d$ and their standard deviations, summarized in Table \ref{Tab-1} ($\rho=0$), Table \ref{Tab-2} ($\rho=0.4$), and Table \ref{Tab-3} ($\rho=0.8$).

\begin{table}[htbp]
  \caption
  {Prediction performances of ICSM, ICM and ISM when $\rho=0$.}
  \label{Tab-1}
  \setlength\tabcolsep{4.8pt}
  \renewcommand{\arraystretch}{1.2}
  \begin{center}
  \scriptsize
    \begin{tabular}{ccccccccccc}
    \hline   \specialrule{0.05em}{1pt}{1pt}
 &&&\multicolumn{3}{c}{$MSE_l$}&&\multicolumn{3}{c}{$MSE_u$}\\
 \cline{4-6}\cline{8-10}
n&$\epsilon_i^c$& $\bm{W}$ & ICSM & ICM & ISM & & ICSM & ICM & ISM \\
\hline
120	&	N(0,11)	&	rook	&$\underset{(	2.0523	)}{	11.1014	}$&$\underset{(	2.0225	)}{	11.0722	}$&$\underset{(	2.1043	)}{	11.1052	}$&&$\underset{(	2.2172	)}{	11.2156	}$&$\underset{(	2.1824	)}{	11.1783	}$&$\underset{(	2.2575	)}{	11.2237	}$\\
	&		&	block	&$\underset{(	2.1504	)}{	11.2169	}$&$\underset{(	2.1351	)}{	11.1144	}$&$\underset{(	2.1985	)}{	11.2134	}$&&$\underset{(	2.3606	)}{	11.2167	}$&$\underset{(	2.3259	)}{	11.0920	}$&$\underset{(	2.4171	)}{	11.2161	}$\\
\cline{4-10}																													
	&	N(0,18)	&	rook	&$\underset{(	3.5658	)}{	18.4828	}$&$\underset{(	3.5652	)}{	18.3768	}$&$\underset{(	3.5224	)}{	18.4603	}$&&$\underset{(	3.4763	)}{	18.8141	}$&$\underset{(	3.4616	)}{	18.7145	}$&$\underset{(	3.4930	)}{	18.5812	}$\\
	&		&	block	&$\underset{(	3.7560	)}{	18.4331	}$&$\underset{(	3.7507	)}{	18.4424	}$&$\underset{(	3.6357	)}{	18.4669	}$&&$\underset{(	3.7473	)}{	18.3173	}$&$\underset{(	3.7632	)}{	18.3298	}$&$\underset{(	3.6742	)}{	18.3754	}$\\
\hline																													
240	&	N(0,11)	&	rook	&$\underset{(	1.6740	)}{	11.0143	}$&$\underset{(	1.6876	)}{	11.0286	}$&$\underset{(	1.6674	)}{	11.0579	}$&&$\underset{(	1.6481	)}{	11.1054	}$&$\underset{(	1.6549	)}{	11.1150	}$&$\underset{(	1.6442	)}{	11.1523	}$\\
	&		&	block	&$\underset{(	1.4210	)}{	11.2274	}$&$\underset{(	1.4090	)}{	11.2125	}$&$\underset{(	1.4332	)}{	11.2541	}$&&$\underset{(	1.4140	)}{	11.1916	}$&$\underset{(	1.4049	)}{	11.1743	}$&$\underset{(	1.4258	)}{	11.2184	}$\\
\cline{4-10}																													
	&	N(0,18)	&	rook	&$\underset{(	2.6180	)}{	18.1153	}$&$\underset{(	2.6334	)}{	18.1168	}$&$\underset{(	2.5155	)}{	17.9986	}$&&$\underset{(	2.4520	)}{	17.9500	}$&$\underset{(	2.4717	)}{	17.9420	}$&$\underset{(	2.4846	)}{	17.9134	}$\\
	&		&	block	&$\underset{(	3.2038	)}{	17.9657	}$&$\underset{(	3.1798	)}{	17.9348	}$&$\underset{(	3.0554	)}{	17.7945	}$&&$\underset{(	3.2844	)}{	17.9281	}$&$\underset{(	3.2506	)}{	17.9095	}$&$\underset{(	3.1371	)}{	17.8560	}$\\
\hline																													
500	&	N(0,11)	&	rook	&$\underset{(	0.9895	)}{	10.9784	}$&$\underset{(	0.9850	)}{	10.9739	}$&$\underset{(	0.9851	)}{	10.9848	}$&&$\underset{(	0.9870	)}{	11.0429	}$&$\underset{(	0.9821	)}{	11.0348	}$&$\underset{(	0.9816	)}{	11.0498	}$\\
	&		&	block	&$\underset{(	1.1267	)}{	10.9696	}$&$\underset{(	1.1224	)}{	10.9653	}$&$\underset{(	1.1257	)}{	10.9791	}$&&$\underset{(	1.1408	)}{	11.0228	}$&$\underset{(	1.1370	)}{	11.0189	}$&$\underset{(	1.1387	)}{	11.0335	}$\\
\cline{4-10}																													
	&	N(0,18)	&	rook	&$\underset{(	1.7410	)}{	18.3842	}$&$\underset{(	1.7629	)}{	18.3787	}$&$\underset{(	1.6961	)}{	18.1039	}$&&$\underset{(	1.7857	)}{	18.1269	}$&$\underset{(	1.7940	)}{	18.0955	}$&$\underset{(	1.6838	)}{	18.0033	}$\\
	&		&	block	&$\underset{(	1.9952	)}{	18.5208	}$&$\underset{(	2.0078	)}{	18.5262	}$&$\underset{(	1.8678	)}{	18.2646	}$&&$\underset{(	2.1360	)}{	18.4309	}$&$\underset{(	2.1583	)}{	18.4353	}$&$\underset{(	1.9304	)}{	18.3191	}$\\
\hline
  \specialrule{0.05em}{20pt}{1pt}
  \hline
 &&&\multicolumn{3}{c}{$AR$}&&\multicolumn{3}{c}{$N_d$}\\
 \cline{4-6}\cline{8-10}
n&$\epsilon_i^c$& $\bm{W}$ & ICSM & ICM & ISM & & ICSM & ICM & ISM  \\
\hline
120	&	N(0,11)	&	rook	&$\underset{(	0.0452	)}{	0.7749	}$&$\underset{(	0.0442	)}{	0.7748	}$&$\underset{(	0.0455	)}{	0.7747	}$&&$\underset{(	0.0000	)}{	0.0000	}$&$\underset{(	0.0000	)}{	0.0000	}$&$\underset{(	0.0000	)}{	0.0000	}$\\
	&		&	block	&$\underset{(	0.0441	)}{	0.7725	}$&$\underset{(	0.0435	)}{	0.7743	}$&$\underset{(	0.0448	)}{	0.7727	}$&&$\underset{(	0.0000	)}{	0.0000	}$&$\underset{(	0.0000	)}{	0.0000	}$&$\underset{(	0.0000	)}{	0.0000	}$\\
\cline{4-10}																													
	&	N(0,18)	&	rook	&$\underset{(	0.0726	)}{	0.6453	}$&$\underset{(	0.0724	)}{	0.6467	}$&$\underset{(	0.0732	)}{	0.6446	}$&&$\underset{(	0.2262	)}{	0.0533	}$&$\underset{(	0.2262	)}{	0.0533	}$&$\underset{(	0.4903	)}{	0.6133	}$\\
	&		&	block	&$\underset{(	0.0687	)}{	0.6709	}$&$\underset{(	0.0691	)}{	0.6704	}$&$\underset{(	0.0682	)}{	0.6706	}$&&$\underset{(	0.1155	)}{	0.0133	}$&$\underset{(	0.1155	)}{	0.0133	}$&$\underset{(	0.4957	)}{	0.5867	}$\\
\hline																													
240	&	N(0,11)	&	rook	&$\underset{(	0.0388	)}{	0.7771	}$&$\underset{(	0.0390	)}{	0.7768	}$&$\underset{(	0.0386	)}{	0.7765	}$&&$\underset{(	0.0000	)}{	0.0000	}$&$\underset{(	0.0000	)}{	0.0000	}$&$\underset{(	0.0000	)}{	0.0000	}$\\
	&		&	block	&$\underset{(	0.0373	)}{	0.7693	}$&$\underset{(	0.0373	)}{	0.7695	}$&$\underset{(	0.0374	)}{	0.7687	}$&&$\underset{(	0.0000	)}{	0.0000	}$&$\underset{(	0.0000	)}{	0.0000	}$&$\underset{(	0.0000	)}{	0.0000	}$\\
\cline{4-10}																													
	&	N(0,18)	&	rook	&$\underset{(	0.0541	)}{	0.6672	}$&$\underset{(	0.0538	)}{	0.6671	}$&$\underset{(	0.0552	)}{	0.6666	}$&&$\underset{(	0.2731	)}{	0.0800	}$&$\underset{(	0.2731	)}{	0.0800	}$&$\underset{(	0.1622	)}{	0.9733	}$\\
	&		&	block	&$\underset{(	0.0654	)}{	0.6696	}$&$\underset{(	0.0649	)}{	0.6703	}$&$\underset{(	0.0643	)}{	0.6703	}$&&$\underset{(	0.0000	)}{	0.0000	}$&$\underset{(	0.0000	)}{	0.0000	}$&$\underset{(	0.4124	)}{	0.2133	}$\\
\hline																													
500	&	N(0,11)	&	rook	&$\underset{(	0.0344	)}{	0.7755	}$&$\underset{(	0.0342	)}{	0.7756	}$&$\underset{(	0.0343	)}{	0.7755	}$&&$\underset{(	0.0000	)}{	0.0000	}$&$\underset{(	0.0000	)}{	0.0000	}$&$\underset{(	0.0000	)}{	0.0000	}$\\
	&		&	block	&$\underset{(	0.0326	)}{	0.7792	}$&$\underset{(	0.0325	)}{	0.7793	}$&$\underset{(	0.0325	)}{	0.7791	}$&&$\underset{(	0.0000	)}{	0.0000	}$&$\underset{(	0.0000	)}{	0.0000	}$&$\underset{(	0.0000	)}{	0.0000	}$\\
\cline{4-10}																													
	&	N(0,18)	&	rook	&$\underset{(	0.0395	)}{	0.6711	}$&$\underset{(	0.0399	)}{	0.6712	}$&$\underset{(	0.0402	)}{	0.6719	}$&&$\underset{(	0.2731	)}{	0.0800	}$&$\underset{(	0.2731	)}{	0.0800	}$&$\underset{(	0.0000	)}{	1.0000	}$\\
	&		&	block	&$\underset{(	0.0469	)}{	0.6668	}$&$\underset{(	0.0470	)}{	0.6667	}$&$\underset{(	0.0467	)}{	0.6666	}$&&$\underset{(	0.3661	)}{	0.1200	}$&$\underset{(	0.2929	)}{	0.0933	}$&$\underset{(	0.7884	)}{	1.2000	}$\\

\hline
    \end{tabular}
  \end{center}
\end{table}

\begin{table}[htbp]
  \caption
  {Prediction performances of ICSM, ICM and ISM when $\rho=0.4$.}
  \label{Tab-2}
  \setlength\tabcolsep{4.8pt}
  \renewcommand{\arraystretch}{1.2}
  \begin{center}
  \scriptsize
    \begin{tabular}{ccccccccccc}
    \hline  \specialrule{0.05em}{1pt}{1pt}
 &&&\multicolumn{3}{c}{$MSE_l$}&&\multicolumn{3}{c}{$MSE_u$}\\
 \cline{4-6}\cline{8-10}
n&$\epsilon_i^c$& $\bm{W}$ & ICSM & ICM & ISM & & ICSM & ICM & ISM \\
\hline
120	&	N(0,11)	&	rook	&$\underset{(	2.2722	)}{	11.4445	}$&$\underset{(	6.8898	)}{	26.8926	}$&$\underset{(	2.1702	)}{	11.4144	}$&&$\underset{(	2.1354	)}{	11.4818	}$&$\underset{(	7.8028	)}{	27.2150	}$&$\underset{(	2.1127	)}{	11.4899	}$\\
	&		&	block	&$\underset{(	2.3685	)}{	10.8221	}$&$\underset{(	6.7069	)}{	29.6401	}$&$\underset{(	2.3716	)}{	10.8411	}$&&$\underset{(	2.1260	)}{	10.8115	}$&$\underset{(	6.1561	)}{	29.3916	}$&$\underset{(	2.1271	)}{	10.8306	}$\\
\cline{4-10}																													
	&	N(0,18)	&	rook	&$\underset{(	3.9168	)}{	18.1000	}$&$\underset{(	8.0248	)}{	32.3855	}$&$\underset{(	3.8590	)}{	18.1169	}$&&$\underset{(	4.2566	)}{	18.3430	}$&$\underset{(	6.9920	)}{	31.0777	}$&$\underset{(	3.6878	)}{	18.1409	}$\\
	&		&	block	&$\underset{(	3.2555	)}{	17.4380	}$&$\underset{(	8.8245	)}{	34.4353	}$&$\underset{(	3.2768	)}{	17.4085	}$&&$\underset{(	3.2531	)}{	17.4482	}$&$\underset{(	7.6091	)}{	33.6160	}$&$\underset{(	3.2425	)}{	17.3586	}$\\
\hline																													
240	&	N(0,11)	&	rook	&$\underset{(	1.5169	)}{	11.2280	}$&$\underset{(	4.8289	)}{	26.5330	}$&$\underset{(	1.5527	)}{	11.2390	}$&&$\underset{(	1.5245	)}{	11.2742	}$&$\underset{(	5.2139	)}{	26.6516	}$&$\underset{(	1.5748	)}{	11.2438	}$\\
	&		&	block	&$\underset{(	1.6397	)}{	11.0554	}$&$\underset{(	4.5317	)}{	22.5821	}$&$\underset{(	1.6432	)}{	11.0636	}$&&$\underset{(	1.6779	)}{	10.9353	}$&$\underset{(	4.0878	)}{	22.0217	}$&$\underset{(	1.6815	)}{	10.9421	}$\\
\cline{4-10}																													
	&	N(0,18)	&	rook	&$\underset{(	3.0075	)}{	17.9222	}$&$\underset{(	7.0524	)}{	31.9797	}$&$\underset{(	2.6908	)}{	17.6607	}$&&$\underset{(	3.0995	)}{	18.0094	}$&$\underset{(	6.6212	)}{	31.8204	}$&$\underset{(	2.6683	)}{	17.6132	}$\\
	&		&	block	&$\underset{(	2.4087	)}{	17.6641	}$&$\underset{(	5.3542	)}{	27.5415	}$&$\underset{(	2.4046	)}{	17.5935	}$&&$\underset{(	2.3808	)}{	17.6786	}$&$\underset{(	4.1114	)}{	26.2599	}$&$\underset{(	2.3896	)}{	17.5969	}$\\
\hline																													
500	&	N(0,11)	&	rook	&$\underset{(	1.2764	)}{	11.4305	}$&$\underset{(	5.0379	)}{	28.5840	}$&$\underset{(	1.0681	)}{	11.2410	}$&&$\underset{(	1.7726	)}{	11.5519	}$&$\underset{(	4.5871	)}{	28.0278	}$&$\underset{(	1.0960	)}{	11.2111	}$\\
	&		&	block	&$\underset{(	1.1075	)}{	11.1313	}$&$\underset{(	2.4013	)}{	18.3621	}$&$\underset{(	1.1116	)}{	11.1339	}$&&$\underset{(	1.1124	)}{	11.0830	}$&$\underset{(	2.4578	)}{	18.3696	}$&$\underset{(	1.1171	)}{	11.0864	}$\\
\cline{4-10}																													
	&	N(0,18)	&	rook	&$\underset{(	2.5031	)}{	18.7708	}$&$\underset{(	5.7933	)}{	33.8118	}$&$\underset{(	2.0405	)}{	18.0857	}$&&$\underset{(	2.1667	)}{	18.3586	}$&$\underset{(	5.9500	)}{	34.6654	}$&$\underset{(	2.0512	)}{	18.0755	}$\\
	&		&	block	&$\underset{(	2.3434	)}{	18.2823	}$&$\underset{(	4.3289	)}{	24.8806	}$&$\underset{(	2.2057	)}{	18.0818	}$&&$\underset{(	2.1289	)}{	18.1978	}$&$\underset{(	3.1235	)}{	24.6113	}$&$\underset{(	2.1457	)}{	18.0764	}$\\

\hline
  \specialrule{0.05em}{20pt}{1pt}
\hline
 &&&\multicolumn{3}{c}{$AR$}&&\multicolumn{3}{c}{$N_d$}\\
 \cline{4-6}\cline{8-10}
n&$\epsilon_i^c$& $\bm{W}$ & ICSM & ICM & ISM & & ICSM & ICM & ISM  \\
\hline
120	&	N(0,11)	&	rook	&$\underset{(	0.0526	)}{	0.7697	}$&$\underset{(	0.0994	)}{	0.5660	}$&$\underset{(	0.0521	)}{	0.7695	}$&&$\underset{(	0.0000	)}{	0.0000	}$&$\underset{(	0.4124	)}{	0.2133	}$&$\underset{(	0.0000	)}{	0.0000	}$\\
	&		&	block	&$\underset{(	0.0510	)}{	0.7795	}$&$\underset{(	0.0835	)}{	0.5363	}$&$\underset{(	0.0513	)}{	0.7791	}$&&$\underset{(	0.0000	)}{	0.0000	}$&$\underset{(	0.3422	)}{	0.1333	}$&$\underset{(	0.0000	)}{	0.0000	}$\\
\cline{4-10}																													
	&	N(0,18)	&	rook	&$\underset{(	0.0719	)}{	0.6658	}$&$\underset{(	0.0836	)}{	0.5214	}$&$\underset{(	0.0720	)}{	0.6640	}$&&$\underset{(	0.2262	)}{	0.0533	}$&$\underset{(	0.5012	)}{	0.2133	}$&$\underset{(	0.2929	)}{	0.9067	}$\\
	&		&	block	&$\underset{(	0.0600	)}{	0.6761	}$&$\underset{(	0.0910	)}{	0.5075	}$&$\underset{(	0.0608	)}{	0.6769	}$&&$\underset{(	0.0000	)}{	0.0000	}$&$\underset{(	0.4378	)}{	0.2533	}$&$\underset{(	0.0000	)}{	0.0000	}$\\
\hline																													
240	&	N(0,11)	&	rook	&$\underset{(	0.0406	)}{	0.7738	}$&$\underset{(	0.0694	)}{	0.5728	}$&$\underset{(	0.0411	)}{	0.7737	}$&&$\underset{(	0.0000	)}{	0.0000	}$&$\underset{(	0.3661	)}{	0.1200	}$&$\underset{(	0.0000	)}{	0.0000	}$\\
	&		&	block	&$\underset{(	0.0402	)}{	0.7739	}$&$\underset{(	0.0701	)}{	0.6112	}$&$\underset{(	0.0402	)}{	0.7738	}$&&$\underset{(	0.0000	)}{	0.0000	}$&$\underset{(	0.3108	)}{	0.1067	}$&$\underset{(	0.0000	)}{	0.0000	}$\\
\cline{4-10}																													
	&	N(0,18)	&	rook	&$\underset{(	0.0582	)}{	0.6755	}$&$\underset{(	0.0709	)}{	0.5365	}$&$\underset{(	0.0560	)}{	0.6780	}$&&$\underset{(	0.2262	)}{	0.0533	}$&$\underset{(	0.6905	)}{	0.3600	}$&$\underset{(	0.1622	)}{	0.9733	}$\\
	&		&	block	&$\underset{(	0.0513	)}{	0.6766	}$&$\underset{(	0.0655	)}{	0.5674	}$&$\underset{(	0.0522	)}{	0.6761	}$&&$\underset{(	0.1155	)}{	0.0133	}$&$\underset{(	0.5774	)}{	0.2667	}$&$\underset{(	0.3691	)}{	0.8400	}$\\
\hline																													
500	&	N(0,11)	&	rook	&$\underset{(	0.0406	)}{	0.7699	}$&$\underset{(	0.0581	)}{	0.5633	}$&$\underset{(	0.0374	)}{	0.7736	}$&&$\underset{(	0.0000	)}{	0.0000	}$&$\underset{(	0.5733	)}{	0.3200	}$&$\underset{(	0.0000	)}{	0.0000	}$\\
	&		&	block	&$\underset{(	0.0355	)}{	0.7732	}$&$\underset{(	0.0478	)}{	0.6658	}$&$\underset{(	0.0355	)}{	0.7732	}$&&$\underset{(	0.0000	)}{	0.0000	}$&$\underset{(	0.3002	)}{	0.0667	}$&$\underset{(	0.0000	)}{	0.0000	}$\\
\cline{4-10}																													
	&	N(0,18)	&	rook	&$\underset{(	0.0467	)}{	0.6700	}$&$\underset{(	0.0653	)}{	0.5252	}$&$\underset{(	0.0463	)}{	0.6707	}$&&$\underset{(	0.2929	)}{	0.0933	}$&$\underset{(	0.6595	)}{	0.4133	}$&$\underset{(	0.4027	)}{	1.0000	}$\\
	&		&	block	&$\underset{(	0.0494	)}{	0.6670	}$&$\underset{(	0.0576	)}{	0.5922	}$&$\underset{(	0.0503	)}{	0.6673	}$&&$\underset{(	0.3188	)}{	0.0800	}$&$\underset{(	0.5199	)}{	0.2000	}$&$\underset{(	1.0342	)}{	1.2267	}$\\

\hline
    \end{tabular}
  \end{center}
\end{table}

\begin{table}[htbp]
  \caption
  {Prediction performances of ICSM, ICM and ISM when $\rho=0.8$.}
  \label{Tab-3}
  \setlength\tabcolsep{4.8pt}
  \renewcommand{\arraystretch}{1.2}
  \begin{center}
  \scriptsize
    \begin{tabular}{ccccccccccc}
    \hline   \specialrule{0.05em}{1pt}{1pt}
 &&&\multicolumn{3}{c}{$MSE_l$}&&\multicolumn{3}{c}{$MSE_u$}\\
 \cline{4-6}\cline{8-10}
n&$\epsilon_i^c$& $\bm{W}$ & ICSM & ICM & ISM & & ICSM & ICM & ISM \\
\hline
120	&	N(0,11)	&	rook	&$\underset{(	9.6780	)}{	20.9170	}$&$\underset{(	46.9563	)}{	136.9090	}$&$\underset{(	2.6969	)}{	11.8538	}$&&$\underset{(	13.6396	)}{	22.1827	}$&$\underset{(	43.4317	)}{	143.8223	}$&$\underset{(	2.7121	)}{	11.9363	}$\\
	&		&	block	&$\underset{(	2.1735	)}{	10.0668	}$&$\underset{(	74.8370	)}{	268.4357	}$&$\underset{(	2.1649	)}{	10.0871	}$&&$\underset{(	2.1566	)}{	10.1485	}$&$\underset{(	67.7378	)}{	276.4214	}$&$\underset{(	2.1272	)}{	10.1281	}$\\
\cline{4-10}																													
	&	N(0,18)	&	rook	&$\underset{(	11.7355	)}{	26.8112	}$&$\underset{(	38.1187	)}{	138.2213	}$&$\underset{(	4.1060	)}{	18.2134	}$&&$\underset{(	11.7760	)}{	25.2355	}$&$\underset{(	44.4695	)}{	135.1789	}$&$\underset{(	3.8650	)}{	18.2767	}$\\
	&		&	block	&$\underset{(	3.8754	)}{	17.7788	}$&$\underset{(	75.3455	)}{	276.9172	}$&$\underset{(	3.5006	)}{	17.5684	}$&&$\underset{(	3.8585	)}{	18.1281	}$&$\underset{(	82.8051	)}{	286.9323	}$&$\underset{(	3.3895	)}{	17.6841	}$\\
\hline																													
240	&	N(0,11)	&	rook	&$\underset{(	12.8142	)}{	22.5784	}$&$\underset{(	33.9702	)}{	145.8142	}$&$\underset{(	1.8895	)}{	11.5306	}$&&$\underset{(	11.8816	)}{	22.3133	}$&$\underset{(	37.3522	)}{	155.3351	}$&$\underset{(	1.8967	)}{	11.4477	}$\\
	&		&	block	&$\underset{(	1.5862	)}{	10.9630	}$&$\underset{(	53.9260	)}{	193.0556	}$&$\underset{(	1.5832	)}{	10.9685	}$&&$\underset{(	1.5722	)}{	10.7834	}$&$\underset{(	51.5894	)}{	190.3062	}$&$\underset{(	1.5694	)}{	10.7810	}$\\
\cline{4-10}																													
	&	N(0,18)	&	rook	&$\underset{(	10.9487	)}{	29.6531	}$&$\underset{(	40.5606	)}{	167.7000	}$&$\underset{(	2.5728	)}{	17.4381	}$&&$\underset{(	12.7241	)}{	32.8813	}$&$\underset{(	44.8410	)}{	165.4525	}$&$\underset{(	2.6281	)}{	17.4703	}$\\
	&		&	block	&$\underset{(	2.9300	)}{	18.1834	}$&$\underset{(	43.7782	)}{	195.6656	}$&$\underset{(	2.6648	)}{	17.8889	}$&&$\underset{(	2.6466	)}{	18.1563	}$&$\underset{(	56.4152	)}{	202.0845	}$&$\underset{(	2.5530	)}{	17.8429	}$\\
\hline																													
500	&	N(0,11)	&	rook	&$\underset{(	14.5872	)}{	27.1713	}$&$\underset{(	37.4465	)}{	173.1929	}$&$\underset{(	1.1855	)}{	11.1199	}$&&$\underset{(	13.2115	)}{	25.5140	}$&$\underset{(	33.4171	)}{	167.6693	}$&$\underset{(	1.1698	)}{	11.0789	}$\\
	&		&	block	&$\underset{(	1.1800	)}{	11.0527	}$&$\underset{(	36.9454	)}{	147.2848	}$&$\underset{(	1.1786	)}{	11.0548	}$&&$\underset{(	1.1966	)}{	11.1206	}$&$\underset{(	35.1845	)}{	142.7372	}$&$\underset{(	1.1941	)}{	11.1228	}$\\
\cline{4-10}																													
	&	N(0,18)	&	rook	&$\underset{(	10.7293	)}{	31.5665	}$&$\underset{(	35.4408	)}{	180.5881	}$&$\underset{(	1.6784	)}{	17.4332	}$&&$\underset{(	12.5326	)}{	32.3119	}$&$\underset{(	32.0515	)}{	179.2781	}$&$\underset{(	1.6377	)}{	17.4337	}$\\
	&		&	block	&$\underset{(	2.1603	)}{	18.2320	}$&$\underset{(	47.6102	)}{	163.7546	}$&$\underset{(	1.7004	)}{	17.7877	}$&&$\underset{(	1.9119	)}{	17.9188	}$&$\underset{(	45.7619	)}{	161.2984	}$&$\underset{(	1.7644	)}{	17.8166	}$\\
\hline
  \specialrule{0.05em}{20pt}{1pt}
  \hline
 &&&\multicolumn{3}{c}{$AR$}&&\multicolumn{3}{c}{$N_d$}\\
 \cline{4-6}\cline{8-10}
n&$\epsilon_i^c$& $\bm{W}$ & ICSM & ICM & ISM & & ICSM & ICM & ISM  \\
\hline
120	&	N(0,11)	&	rook	&$\underset{(	0.1076	)}{	0.6513	}$&$\underset{(	0.0548	)}{	0.2094	}$&$\underset{(	0.0627	)}{	0.7540	}$&&$\underset{(	0.0000	)}{	0.0000	}$&$\underset{(	0.6993	)}{	0.5867	}$&$\underset{(	0.0000	)}{	0.0000	}$\\
	&		&	block	&$\underset{(	0.0503	)}{	0.7910	}$&$\underset{(	0.0314	)}{	0.1222	}$&$\underset{(	0.0502	)}{	0.7911	}$&&$\underset{(	0.0000	)}{	0.0000	}$&$\underset{(	0.8246	)}{	0.6800	}$&$\underset{(	0.0000	)}{	0.0000	}$\\
\cline{4-10}																													
	&	N(0,18)	&	rook	&$\underset{(	0.0991	)}{	0.6168	}$&$\underset{(	0.0586	)}{	0.2234	}$&$\underset{(	0.0672	)}{	0.6769	}$&&$\underset{(	0.1622	)}{	0.0267	}$&$\underset{(	0.8904	)}{	0.7333	}$&$\underset{(	0.4520	)}{	0.2800	}$\\
	&		&	block	&$\underset{(	0.0683	)}{	0.6611	}$&$\underset{(	0.0331	)}{	0.1168	}$&$\underset{(	0.0692	)}{	0.6590	}$&&$\underset{(	0.1155	)}{	0.0133	}$&$\underset{(	0.6443	)}{	0.5200	}$&$\underset{(	0.5030	)}{	0.4800	}$\\
\hline																													
240	&	N(0,11)	&	rook	&$\underset{(	0.1138	)}{	0.6641	}$&$\underset{(	0.0444	)}{	0.2143	}$&$\underset{(	0.0406	)}{	0.7792	}$&&$\underset{(	0.0000	)}{	0.0000	}$&$\underset{(	0.8268	)}{	0.5467	}$&$\underset{(	0.0000	)}{	0.0000	}$\\
	&		&	block	&$\underset{(	0.0387	)}{	0.7811	}$&$\underset{(	0.0415	)}{	0.1660	}$&$\underset{(	0.0387	)}{	0.7811	}$&&$\underset{(	0.0000	)}{	0.0000	}$&$\underset{(	0.7565	)}{	0.5733	}$&$\underset{(	0.0000	)}{	0.0000	}$\\
\cline{4-10}																													
	&	N(0,18)	&	rook	&$\underset{(	0.0914	)}{	0.5594	}$&$\underset{(	0.0433	)}{	0.1860	}$&$\underset{(	0.0558	)}{	0.6680	}$&&$\underset{(	0.0000	)}{	0.0000	}$&$\underset{(	0.7236	)}{	0.4933	}$&$\underset{(	0.4452	)}{	0.7333	}$\\
	&		&	block	&$\underset{(	0.0508	)}{	0.6730	}$&$\underset{(	0.0369	)}{	0.1602	}$&$\underset{(	0.0506	)}{	0.6748	}$&&$\underset{(	0.1622	)}{	0.0267	}$&$\underset{(	0.7739	)}{	0.6800	}$&$\underset{(	0.5030	)}{	0.5200	}$\\
\hline																													
500	&	N(0,11)	&	rook	&$\underset{(	0.1180	)}{	0.6192	}$&$\underset{(	0.0425	)}{	0.1867	}$&$\underset{(	0.0376	)}{	0.7738	}$&&$\underset{(	0.0000	)}{	0.0000	}$&$\underset{(	0.6844	)}{	0.4667	}$&$\underset{(	0.0000	)}{	0.0000	}$\\
	&		&	block	&$\underset{(	0.0357	)}{	0.7718	}$&$\underset{(	0.0439	)}{	0.2038	}$&$\underset{(	0.0357	)}{	0.7717	}$&&$\underset{(	0.0000	)}{	0.0000	}$&$\underset{(	0.6611	)}{	0.5733	}$&$\underset{(	0.0000	)}{	0.0000	}$\\
\cline{4-10}																													
	&	N(0,18)	&	rook	&$\underset{(	0.0935	)}{	0.5617	}$&$\underset{(	0.0325	)}{	0.1746	}$&$\underset{(	0.0389	)}{	0.6768	}$&&$\underset{(	0.1155	)}{	0.0133	}$&$\underset{(	0.7948	)}{	0.5067	}$&$\underset{(	0.2596	)}{	0.9867	}$\\
	&		&	block	&$\underset{(	0.0462	)}{	0.6665	}$&$\underset{(	0.0468	)}{	0.1881	}$&$\underset{(	0.0456	)}{	0.6670	}$&&$\underset{(	0.4638	)}{	0.1200	}$&$\underset{(	0.7493	)}{	0.6267	}$&$\underset{(	0.6513	)}{	1.1867	}$\\
\hline
    \end{tabular}
  \end{center}
\end{table}

\section{Application}\label{sec5}
In this section, we give a detailed analysis of the weather data mentioned in Section \ref{sec1} using the proposed new model. In particular, to examine the prediction power of the ICSM, we collected monthly temperature and monthly rainfall data in the year of $2020$ from China Meteorological Yearbook. That is, data in $2019$ acts as training set, and those in $2020$ acts as test set. In the preprocessing of the data, we first take logarithm of the precipitation, then group all the single values into intervals. The ICSM is built as
\begin{equation*}\label{mod-5}
\left\{
\begin{array}{ll}
y_i^c = \rho \sum_{i^\prime\neq i}w_{ii^\prime}y_{i^\prime} + x_i^c\beta_1^c+x_i^r\beta_2^c+\epsilon_i^c  \\
  y_i^r = x_i^c\beta_1^r+x_i^r\beta_2^r+\epsilon_i^r
\end{array}
\right.
\end{equation*}
\begin{equation*}
 s.t.  \quad \left\{\begin{array}{ll}
y_i^c-\rho \sum_{i^\prime\neq i}w_{ii^\prime}y_{i^\prime}-y_i^r  \leq x_i^c\beta_1^c+x_i^r\beta_2^c+x_i^c\beta_1^r+x_i^r\beta_2^r\\
y_i^c-\rho \sum_{i^\prime\neq i}w_{ii^\prime}y_{i^\prime}+y_i^r  \geq x_i^c\beta_1^c+x_i^r\beta_2^c-(x_i^c\beta_1^r+x_i^r\beta_2^r)\\
x_i^c\beta_1^r+x_i^r\beta_2^r \geq 0 .\\
\end{array}\right.
\end{equation*}
And the ICM is
\begin{equation}\label{cscrm}
\left\{
\begin{array}{ll}
y_i^c =  x_i^c\beta_1^c+x_i^r\beta_2^c+\epsilon_i^c  \\
  y_i^r = x_i^c\beta_1^r+x_i^r\beta_2^r+\epsilon_i^r
\end{array}
\right.
 \quad  s.t.  \quad \left\{\begin{array}{ll}
y_i^c-y_i^r  \leq x_i^c\beta_1^c+x_i^r\beta_2^c+x_i^c\beta_1^r+x_i^r\beta_2^r\\
y_i^c+y_i^r  \geq x_i^c\beta_1^c+x_i^r\beta_2^c-(x_i^c\beta_1^r+x_i^r\beta_2^r)\\
x_i^c\beta_1^r+x_i^r\beta_2^r \geq 0 .\\
\end{array}\right.
\end{equation}

where $y_i^c, ~y_i^r$ are center and radius of the interval-valued precipitation of city $i$, $x_i^c$ and $x_i^r$ are mid-point and semi-length of the interval-valued temperature of city $i$, $w_{ii^\prime}$ is the spatial weight between cities $i$ and $i^\prime$. $\rho, \beta_1^c,\beta_2^c,\beta_1^r,\beta_2^r$ are parameters to be estimated.

In model \ref{mod-5}, we utilize the inverse distance to obtain $w_{ii^\prime}$, i.e., $w_{ii^\prime}=\frac{1}{d_{ii^\prime}}$, where $d_{ii^\prime}$ is the distance between cities $i$ and $i^\prime$ computed by longitudes and latitudes. Figure \ref{} displays locations of these cities on map of China. To determine a proper $\bm{W}$ for the ICSM, we consider two factors. The first is the number of neighbors $k$, which determines the sparsity of $\bm{W}$. The second is the distance threshold $d_0$. It is known that the greater the $d_{ii^\prime}$, the weaker the spatial effects. Thus we set $w_{ii^\prime}=0$ if $d_{ii^\prime}> d_0$, and $\frac{1}{d_{ii^\prime}}$ otherwise. To choose an optimal pair of $(k, d_0)$ for $\bm{W}$, we firstly set $k=\{1,2,3,\cdots,9\}$. Then regarding each $k$, search for a $d_0$, which makes the Moran's I statistic of $\bm{y}^c$ take the greatest value. Table \ref{Tab.k} shows results for $k$ and $d_0$. It can bee seen that $k=2$ with $d_0=492.37$ km is the best choice.

\begin{table}[!h]
\footnotesize
\caption{ Values of the Moran's I statistics of $y_i^c$s when $k=\{1,2,3,\cdots,9\}$.} \label{Tab.k}
  \setlength\tabcolsep{5pt}
  \renewcommand{\arraystretch}{1.5}
\centering
\begin{tabular}[H]{cccccccccccc}
\specialrule{0.05em}{2pt}{2pt}
   $k$ & $1$ & $2$ & $3$ & $4$ & $5$ & $6$ & $7$ & $8$ & $9$ \\
  \specialrule{0.05em}{2pt}{2pt}
   Moran's I & 0.678 & 0.687  &  0.636 & 0.641 & 0.643  & 0.648 & 0.646 & 0.646 & 0.646 \\ \specialrule{0em}{-.5pt}{-.5pt}
   $d_0$ (km) & 446.57  & 492.37  & 492.37   & 492.37 & 492.37 & 492.37  & 492.37 & 492.37 & 492.37 \\ \specialrule{0em}{-2.5pt}{-2.5pt}
  \specialrule{0.05em}{7pt}{7pt}
\end{tabular}
\end{table}

\begin{figure}[htbp]
    \centering
    \subfigure{
    \resizebox{3.6cm}{3.2cm}{\includegraphics{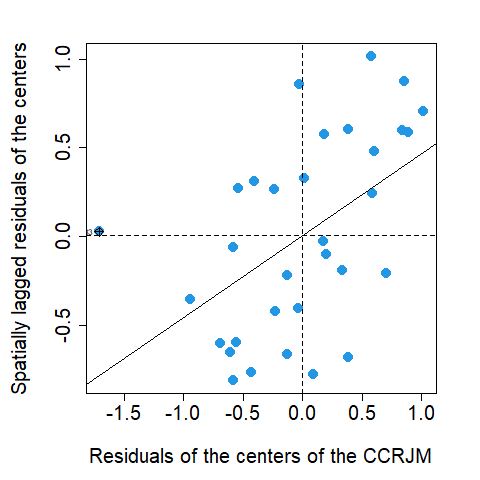}}}
   \subfigure{
  \resizebox{3.6cm}{3.2cm}{\includegraphics{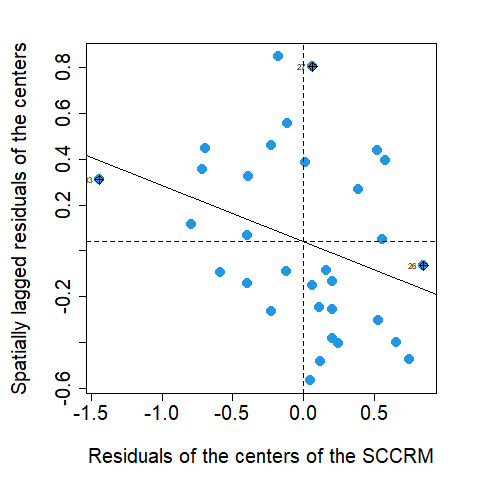}}}
     \subfigure{
  \resizebox{3.6cm}{3.2cm}{\includegraphics{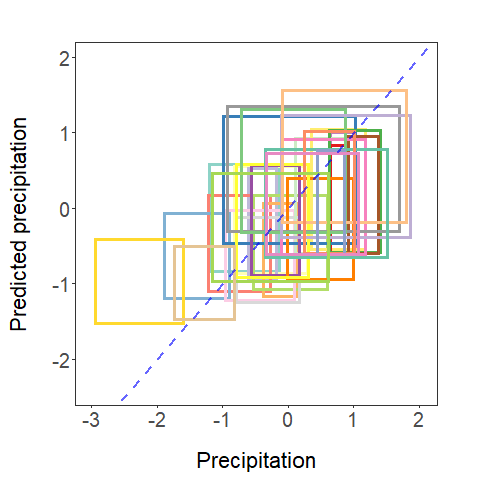}}}
  \caption{The Moran'I Scatter plot of the residuals of $y_i^c$s in the ICM (left) and the ICSM (middle), the predicted $2020$ precipitation intervals versus the true values using the ICSM (right).} \label{fig-1}
\end{figure}

Table \ref{Tab.pre} compares the ICSM and the ICM in terms of both fitting and prediction performances. It is obvious that the ICSM behaves better than the ICM. We can see spatial dependencies are eliminated in the residuals of the ICSM (with the Moran's I statistic being $-0.24$), while still present in those of the ICM (with the Moran's I statistic being $0.46$). From Figure \ref{fig-1} (left and middle) we can come to similar conclusion, which displays Moran's I scatter plots of the residuals. Besides, MSE of the residuals of the ICSM is $0.580$, smaller than that of the ICM ($0.674$). Regarding prediction performance, the ICSM has wider overlapping areas ($44.23\%$) than the ICM ($43.96\%$) by average, and smaller forecast errors of the lower and the upper bounds of the precipitation intervals. Figure \ref{fig-1} (right) exhibits predicted intervals versus the real values.

\begin{table}[!h]
\footnotesize
\caption{Fitting and prediction results of the ICSM and the ICM.}\label{Tab.pre}
  \setlength\tabcolsep{1pt}
  \renewcommand{\arraystretch}{1.5}
\centering
\begin{tabular}[H]{ccccccc}
\specialrule{0.05em}{3pt}{3pt}
   Models &  \parbox[c]{3cm}{Moran's I statistic \newline (residuals of $y_i^c$s)}	&   \parbox[c]{2cm}{MSE  (fitt\newline ed error)} & \parbox[c]{2.5cm}{MSE (predi\newline ction error)}
  & AR & MSE$_l$ & MSE$_u$ \\
  \specialrule{0.05em}{3pt}{3pt}
   ICSM & -0.24 (0.901)& 0.580 & 0.668  & 44.23\% & 0.790 & 0.519  \\
  \specialrule{0em}{-.5pt}{-.5pt}
   ICM  & 0.46 (0.002) &  0.674  & 0.712 & 43.96\% & 0.822 & 0.583  \\
  \specialrule{0em}{-3.5pt}{-3.5pt}
  \specialrule{0.05em}{7pt}{7pt}
\end{tabular}
\end{table}

Table \ref{Tab.fit} summarizes estimation results of the ICSM and the ICM. The spatial lag parameter $\hat{\rho}$ is $0.56$, meaning spatial effects in the precipitation are medium. $\hat{\beta}_1^c$ and $\hat{\beta}_2^c$ are equivalent to $0.261$ and $0.195$, respectively, thus $y_i^c$s are positively related to both centers and ranges of the temperatures. However, $\hat{\beta}_1^r$ equals $0.108$ and $\hat{\beta}_2^r$ is $-0.003$, indicating $y_i^r$s are mainly influenced by mid-points of the temperatures. And semi-lengths of the temperatures play a relatively small role.

\begin{table}[!h]
\footnotesize
\caption{ The fitting results of the ICSM and the ICM.} \label{Tab.fit}
  \setlength\tabcolsep{10.5pt}
  \renewcommand{\arraystretch}{1.5}
\centering
\begin{tabular}[H]{cccccccccc}
\specialrule{0.05em}{3pt}{3pt}
   models & $\hat{\rho}$ & $\hat{\beta}_0^c$ & $\hat{\beta}_1^c$ & $\hat{\beta}_2^c$ & $\hat{\beta}_0^r$ & $\hat{\beta}_1^r$ & $\hat{\beta}_2^r$
  \\
  \specialrule{0.05em}{3pt}{3pt}
   ICSM & 0.56 & -0.167  &  0.261 & 0.195 & 0.710  &  0.108 & -0.003  \\ \specialrule{0em}{-.5pt}{-.5pt}
   ICM &  ---  & -0.250  & 0.510   & 0.284 & 0.710 &   0.108  &  -0.003  \\ \specialrule{0em}{-3.5pt}{-3.5pt}
  \specialrule{0.05em}{7pt}{7pt}
\end{tabular}
\end{table}
\vspace{-0.8cm}

\section{Conclusion}\label{sec6}
Interval-valued data receives much attention due to it's wide applications in the area of meteorology, medicine, finance, etc. Regression models concentrated on this kind of data have also been extended to nonparametric paradigms and the Bayesian framework. Nevertheless, relative few literature considered spatial dependence among intervals. In the article, we put forward a constrained spatial autoregressive model (ICSM) to model interval-valued data with lattice-type spatial correlation. In particular, for the simplicity and efficiency of the new model, we only take spatial effects among centers of responses into account. And to improve prediction accuracy of the model, three inequality constrains for the parameters are added. We obtain the parameters by a grid search technique combined with the constrained least-squares method. Simulation studies show that our new model behaves better than the model without spatial lag term (ICM) and the model without constrains (ISM), when spatial dependencies are present and disturbances are great. And in cases that there are no spatial correlation or disturbances are small, our model performs similarly with the other two models. We also employ a real dataset of temperatures and precipitation to illustrate the utility of the ICSM.

Although we only considered spatial effects among mid-points of outcomes, spatial correlation among the semi-lengths can also be took into consideration by adding a spatial lag term in the linear regression. Under this new model with two different spatial lag parameters, estimator can be gotten by the addition of a ``for'' loop, which searches for an adequate $\rho^r$ for $\bm{y}^r$. Note that the additional iteration would greatly increase the computation time. In future, we will discuss more flexible models, such as the incorporation of two different spatial weight matrices, and more general assumptions of the noise terms.


\bibliography{sn-bibliography}


\end{document}